\documentclass{article}

\begin{document}
\title{Assisted inflation in Bianchi $\textrm{VI}_0$ cosmologies}
\author{
J.~M.~Aguirregabiria$^{1,\,a}$, P.~Labraga$^{1,\,b}$ and Ruth
Lazkoz$^{2,\,c}$\\
\emph{$^1$ Fisika Teorikoaren eta Zientziaren Historiaren Saila}\\
\emph{Euskal Herriko Unibertsitatea,
644 Posta Kutxatila}\\
\emph{ 48080 Bilbao, Spain}\\
\emph{$^2$ Fisika eta Matematika Aplikatuaren Saila}\\
\emph{Deustuko Unibertsitatea,
1 Posta Kutxatila}\\
\emph{ 48080 Bilbao, Spain}\\
\tt{\small $^a$wtpagagj@lg.ehu.es, $^b$wtblalop@lg.ehu.es, $^c$rlazkoz@eside.deusto.es}\\
}

\date{\today}
\maketitle

\begin{abstract}
Exact models for Bianchi $\textrm{VI}_0$ spacetimes with multiple scalar fields 
with exponential potentials have been derived and analysed.
It has been shown that these solutions, when they exist, 
attract neighbouring solutions in the two cases corresponding
to interacting and non-interacting fields. Unlike the results obtained in a previous work
dealing with the late-time inflationary behaviour of Bianchi $\textrm{VI}_0$
cosmologies, the knowledge of exact solutions has made possible 
to study in detail the occurrence of inflation before the asymptotic regime.
As happened in preceding works, here as well inflation
is more likely to happen with a higher number of non-interacting fields
or a lower number of interacting scalar fields.
\end{abstract}

\section{Introduction}

The cosmological inflationary scenario, as pioneerely
proposed by Guth~\cite{Gut81}, has been much favoured due to its 
troubleshooting capabilities~\cite{LidLyt99}. Furthermore, it has recently gained 
the support of observational evidence~\cite{Ber00,Bal00}. It is usually assumed 
that an epoch of accelerated expansion in the Universe was driven by one scalar field
slowly rolling down its potential. Even though  several scalar fields should enter 
the general picture, it is commonplace to make the additional assumption that only 
one of them played a significant dynamical role.

Recently, Liddle, Mazumdar and Schunck~\cite{LidMazSch98} put forward an 
alternative model representing a slight departure from those assumptions.
By means of exact FRW examples, they showed that multiple scalar fields
with exponential potentials can assist each other in the realisation of inflation,
even if the individual fields are not flat enough to produce inflation on their
own. This is the reason why they called this behaviour {\em  assisted inflation}.

The work by Liddle and collaborators was   generalised by Copeland, Mazumdar and 
Nunes~\cite{CopMazNun99} along two different
lines. On the  one hand, they introduced cross-coupling terms in the potential 
and showed that this seems to hinder inflation rather than assist it. On the other 
hand, they analysed cases with multiple exponential potentials and showed
that, by choosing the slopes adequately,  the expansion rate can 
be augmented. Motivation for the use of  exponential potentials is  
can be found in dimensionally reduced supergravity theories 
(see for instance~\cite{CopMazNun99}, \cite{Eas93},
and the references therein).

Assisted inflation presents remarkable novel features from the dynamical point of view.
One would normally expect that a single scalar field 
should eventually dominate the dynamics, but peculiarly inflationary solutions turn
out to be late-time attractors. This fact has been reflected 
in several studies concerned with the dynamics of assisted inflation~\cite{MalWan98,
ColHoo00,LidGre00}.

 Kanti and Olive~\cite{KanOli99b,KanOli99a} explored another interesting aspect 
of the problem
using a realisation of 
the assisted inflation proposal based on the compactification of a 
fifth-dimensional Kaluza-Klein theory. In particular, they showed that assisted inflation 
can be a remedy to the initial conditions problem in the chaotic inflation 
scenario~\cite{Lin83}. Along this line, Liddle and Kaloper~\cite{KalLid00} have found 
that in this kind of inflation the spacetime does retain some memory 
of the  conditions that gave rise to it, even though a homogeneous and
 spatially flat universe is 
generated.

Although most of the references in the literature are concerned with FRW models,
there is no reason to assume that at the onset of inflation the universe
was as isotropic as it seems to be today. In fact, it is commonly believed
that it was precisely an epoch of accelerated
expansion that isotropised the universe. This is the reason why it is of much
relevance to extend the studies on  assisted inflation to Bianchi spacetimes.

Our starting point is a recent collaboration of one of us~\cite{AguChaChi00},  which 
focused on  the study of assisted inflation at late times in FRW and Bianchi I cosmologies
by using exact solutions and in Bianchi $\textrm{VI}_0$
spacetimes by means of asymptotic approximate solutions. 
We construct here a family of exact Bianchi $\textrm{VI}_0$ spacetimes
with exponential potentials which include the exact counterparts of  the mentioned 
approximate solutions. By resorting to these exact, yet particular, solutions we have
 been able to perform a detailed
analysis of the occurrence of inflation in this framework. 

Let us recall that if the models behave as desirable they will undergo 
accelerated expansion 
in their early epochs, preferably for a limited period of time. Clearly, asymptotic 
studies, as those usually done up to the date, have the drawback of being unable to determine 
whether that occurs. 

As will be shown in the following sections, we have found the necessary
and sufficient  conditions to be fulfilled by the slopes of the individual potentials
for the expansion to become accelerated. In the light of the results,
we have been able to illustrate the cumulative effect 
of non-interacting fields in the occurrence of inflation to which assisted
inflation owes its name. What is more, our results exemplify as well the
hindering effect of direct interactions between the fields. For the sake of completeness
we have also included an analysis showing the asymptotic stability of the solutions
as well as an study of the isotropization of our models.
Summarising, our results
confirms previous studies and give additional support to the assisted inflation 
proposal, providing particular exact solutions that attract neighbouring solutions
and show inflation at the start of their life span.
 
\section{Exact solutions to the multiple scalar field problem in 
a Bianchi $\textrm{VI}_0$ spacetime}

In what follows we are going to consider the problem 
of $n$ scalar fields minimally coupled to gravity 
\footnote{Even though most of the papers on the subject deal with 
models in which the fields are minimally coupled to gravity,
inflation can also be assisted in the case of
non-minimally coupled multiple scalar fields with exponential potentials~\cite{Tsu00}.}
  and driven by exponential potentials 
in a Bianchi $\textrm{VI}_0$ spacetime.   First, it will be 
assumed that the scalar fields interact through a product of exponential potentials;
then, we will turn
to consider the alternative case of uncoupled fields evolving
via individual exponential potentials. Nevertheless, it should be pointed out that,
although we will be loosely speaking of interacting and non interacting models,
the geometry will be responsible for some interaction among the 
fields even in the examples devoid of  direct couplings in the potential.

Before we enter the details of the two different cases to be studied, let us
recall the form of the  line element of a Bianchi $\textrm{VI}_0$ spacetime:
\begin{equation}
ds^2=-e^{f(t)}(dt^2-dz^2)+G(t)(e^z dx^2+e^{-z}dy^2).
\end{equation}
For convenience we define two vectors belonging to an $n\textrm{-dimensional}$
Euclidean space as follows:
\begin{eqnarray}
&&\vec\phi=(\phi_1,\phi_2,\dots,\phi_n),\\
&&\vec k=(k_1,k_2,\dots,k_n).
\end{eqnarray}
Moreover, we also demand $\vec k$ be constant with respect to an orthonormal basis
of the space.

\subsection{Non-interacting fields}
In these models we will take 
the potential responsible for the interaction among the fields to be of the form
\begin{equation}
V(\vec \phi)=\sum_{i=1}^nV_{oi}e^{-k_i\phi_i}.
\end{equation}
Motivated by the fact that all scalar fields in these kind of configurations
seem to tend to a common limit \cite{LidMazSch98}, we will make the 
simplifying assumption of having $\phi_1=\phi_2=\dots=\phi_n\equiv\phi$, 
$k_1=k_2=\dots=k_n\equiv k$ and $V_{o1}=V_{o2}=\dots=V_{on}\equiv V_o$.
Additional motivation for this assumption is that it
simplifies considerably the task of finding exact solutions.

It will be possible to determine the model's features upon the resolution of the 
system formed by the Einstein equations
\begin{eqnarray}
&&e^f=\frac{\ddot G}{2\,V\,G},\label{eq_f_2}\\
&&\frac{\ddot G}{G}-\frac{1}{2}\left(\frac{\dot G}{G}\right)^2
-\frac{\dot G}{G}\dot f+\frac{1}{2}=-n\dot{\phi}^2\label{fe4}
\end{eqnarray}
and the Klein-Gordon equation
\begin{equation}
\ddot {\phi}+{\frac{\dot G}{G}}\dot{\phi}-\frac{k}{n}e^f V=0. \label{KEeqbis}
\end{equation}
A first integral to 
(\ref{KEeqbis}) is given by
\begin{equation}
 \dot {\phi}=\frac{k\,\dot G}{2\,n\,G}+\frac{ m}{n\,G} ,
\end{equation}
where $m$ is an integration constant. It is straightforward 
to realise that by means of this first integral and 
eqs.~(\ref{eq_f_2}) and (\ref{fe4}) the problem reduces to finding the solution to 
a single ODE, namely
\begin{equation}
G\ddot G^2-{\buildrel\ldots\over G}\dot G G
+\left(\frac{1}{2}- \frac{k^2 }{4\,n}\right) \ddot G\dot G^2 + \frac{1}{2} \ddot
G G^2+\frac{ m^2}{n} \ddot G = 0 .\label{ode_1}
\end{equation}
Only a particular solution to the case $m=0$ of  eq.~(\ref{ode_1})
has been known to the date~\cite{ChiLab98}.
Remarkably,  it also admits a rather simple family of solutions 
that has not been noticed before.
In order to simplify the expression of these solutions
we introduce a new constant  defined as
\begin{equation}
p= m\,{\frac{\sqrt{ {k^2}-2\,n }}
    {\sqrt{ {k^2}+2\,n} }}\,.
\end{equation}
Up to time-shifts and rescalings of the spatial coordinates $x$ and $y$,
the mentioned solution to eq.~(\ref{ode_1}) is given by
\begin{eqnarray}
\!G(t)=\left(\frac{1}{2}+ p^2\right)\sinh\frac{\sqrt{2n}\,(t+t_0)}
{\sqrt{{k^2}-2n}}+
\left(\frac{1}{2}-
p^2\right)\cosh\frac{\sqrt{2n}\,(t+t_0)}{\sqrt{{k^2}-2n}},\quad
\label{G_general_n}
\end{eqnarray}
and it  exists provided that $k^2>2n$. Note that the solutions 
with $p \ne 0$ have as their late time limit 
the $p=0$ case, the already known solution.


At this point, the complete determination of the remaining metric function $f$ 
and the scalar field $\phi$ is
a straightforward task. Nevertheless, we will omit the 
corresponding  explicit expressions 
because they are  rather lengthy and of little use 
for the forthcoming discussion.

The freedom to choose the value $t_0$ will allows us to avoid unwanted features
in the solutions, in particular we will choose it carefully so that we prevent
signature changes in the $p=0$ cases. To this end we set:
\begin{equation}
t_0=\frac{\sqrt{k^2-2\,n}}{\sqrt{2\,n}}\,\textrm{arctanh}\frac{ 2\,{p^2}-1}{2\,{p^2}+1},
\end{equation}
and thus guarantee $G\ge0$ and $\dot G\ge0$ in the whole range
we are going to consider from now on: $t\ge 0$. Note that no signature changes
arise in the $p=0$ case, so there is in principle no restrictions
upon the  choice of time  origin and we will set $t_0=0$ 
just out of utter convenience. A consequence of these choices is that 
unless $p\ne0$
the curvature scalars will blow up at the beginning of times, i.e. there
will be a spacelike singularity of big-bang type.

\subsection{Interacting fields}
In these other models the potential responsible for the interaction among the fields
will be of the form
\begin{equation}
V(\vec \phi)=\prod_{i=1}^nV_{oi}e^{-k_i\phi_i}.
\end{equation}
The system will then be completely determined upon the resolution of the 
set formed by the Einstein equations
\begin{eqnarray}
&&e^f=\frac{\ddot G}{2\,V\,G},\label{eq_f_1}\\
&&\frac{\ddot G}{G}-\frac{1}{2}\left(\frac{\dot G}{G}\right)^2
-\frac{\dot G}{G}\dot f+\frac{1}{2}=-\dot{\vec\phi}\cdot
\dot {\vec\phi}\label{fe2}
\end{eqnarray}
and the Klein-Gordon equation
\begin{equation}
\ddot {\vec\phi}+{\frac{\dot G}{G}}\dot{\vec\phi}-e^f V\vec k=0 \label{KEequation}.
\end{equation}
One can easily check that the vector
\begin{equation}
 \dot {\vec \phi}=\frac{\dot G}{2\,G}\vec k+\frac{\vec m}{G}
\end{equation}
is a first integral to (\ref{KEequation})
and $\vec m$ another constant vector belonging to the $n$-dimensional Euclidean space,
 that is, 
a set of integration constants. Again, by using this first integral and 
eqs.~(\ref{eq_f_1}) and (\ref{fe2}) the problem reduces to finding the solution to a 
single ODE, 
which does not differ much from the one for the previous case:
\begin{equation}
G\ddot G^2-{\buildrel\ldots\over G}\dot G G
+\left(\frac{1}{2}- \frac{k^2 }{4}\right) \ddot G\dot G^2 + \frac{1}{2} \ddot
G G^2+ m^2 \ddot G = 0 .
\label{ode_2}
\end{equation}
It  follows that the
solutions to eq.~(\ref{ode_2}) can be obtained from those to eq.~(\ref{ode_1})
by remembering that now $k^2=|\vec k|^2$ and $m^2 = \vert\vec m\vert^2$ 
and performing the replacement $n\rightarrow 1$. Correspondingly, 
we introduce a new constant vector defined as
\begin{equation}
\vec p=\vec m\,{\frac{\sqrt{ {k^2}-2  }}
    {\sqrt{ 2 + {k^2}} }}\,.
\end{equation}
We have then
\begin{equation}
G(t)=\left(\frac{1}{2}+p^2\right)
\sinh \frac{\sqrt{2}\,(t+t_0)}{\sqrt{k^2-2}}+
\left(\frac{1}{2}-p^2\right)
\cosh \frac{\sqrt{2}\,(t+t_0)}{\sqrt{k^2-2}},
\label{G_general}
\end{equation}
where we have used  the shorthand $p^2=|\vec p\,|^2$.
Clearly, such solutions exist provided that $k^2>2$. 

The arguments concerning possible signature changes given for the 
models in the previous section apply to these other ones too, so we set
\begin{equation}
t_0=\frac{\sqrt{k^2-2}}{\sqrt{2}}\,\textrm{arctanh}
{\frac{ 2\,{p^2}-1}{2\,{p^2}+1}}
\end{equation}
if $\vec p\ne0$, and $t_0=0$ otherwise.

Summarising, the two families of spacetimes we have constructed are quite 
analogous in their expressions. It remains to see, however, if these similarities
can be extended to their behaviour.
 
\section{Asymptotic behaviour and stability}
The asymptotic behaviour of the kind of ODE leading to our
two families of Bianchi $\textrm{VI}_0$  geometries is worth studying, 
as this analysis 
will enable us to discuss their stability.
In terms of the variable $\Omega$ defined by
\begin{eqnarray}
\Omega=\frac{\dot h}{h^2},\\
h=\frac{\dot G}{G},
\end{eqnarray}
eqs.~(\ref{ode_1}) and (\ref{ode_2}) become
\begin{equation}
\dot \Omega+\left[\Omega+K-\frac{1}{2\,h^2}-\frac{M^2}{\dot G^2}\right]
(\Omega+1)h=0 \label{eq_asym}.
\end{equation}
Here, we have introduced a couple of new parameters defined as 
\begin{equation}
K=\frac{k^2}{4\,n}-\frac{1}{2}
\end{equation}
 and 
\begin{equation}
M^2=\frac{m^2}{n}
\end{equation}
 in the non-interacting case. The corresponding expressions
for the interacting case are obtained by the replacement rules given above.

Equation (\ref{eq_asym}) has the fixed point solution $\Omega_1=-1$. In addition, if 
$\dot G\rightarrow \infty$ asymptotically, then  $\Omega_2=0$ is also  
a fixed point solution. They correspond to $G\propto t$ and 
$G \propto e^{t/\sqrt{2\,K}}$ respectively.
Now, in order to determine whether these two solutions are stable on the $G\rightarrow \infty$
regime we expand each of these solutions about fixed points
by making either $\Omega=\Omega_1+\epsilon$ or $\Omega=\Omega_2+\epsilon$
with $|\epsilon| \ll 1$. For $\Omega_1$ we get
\begin{eqnarray}
\dot \epsilon=\frac{1}{2\,h}\,\epsilon,
\end{eqnarray}
and it indicates that $\Omega_1$ is unstable. In fact, the corresponding solution
$G\propto t$
is spurious because it does not satisfy the Einstein equations (cf.~eqs.~(\ref{eq_f_1}) and 
(\ref{eq_f_2})). 
The case $\Omega_2$ is more interesting because it corresponds precisely to the late-time
approximation
of the exact solutions given above, in this case we get
\begin{eqnarray}
\dot \epsilon=-\frac{1}{\sqrt{2\,K}}\,\epsilon,
\end{eqnarray}
and this proves the  asymptotic stability of our solutions for $K>0$. Therefore, 
for a given $m$ the solution $\Omega_2$ is the attractor for all the 
solutions that are close to it.
\section{Inflation and other kinematic aspects}
It is well known that those spacetimes 
for which the gradient of the scalar field is timelike can be reinterpreted
as perfect fluid induced geometries. By taking advantage 
of this customary reinterpretation we will be able to study the kinematic features
of our spacetimes. For our purposes we just need to know the expression 
of the four velocity of the fluid in terms of the scalar field's gradient, namely
\begin{equation}
u_{\alpha}=\frac{\phi_{,\alpha}}
{\sqrt{-\phi_{,\beta}\phi^{,\beta}}}\qquad\alpha,\beta=0,\dots,3.
\end{equation}
In order to find out whether the model inflates or not it is necessary
to look at the sign of the deceleration parameter 
$q=-\theta^{-2}
\left(3\,\theta_{,\alpha}u^{\alpha}+\theta^2\right)$
where the scalar $\theta=u_{\phantom{\alpha};\alpha}^{\alpha}$
is the  expansion of the fluid~\cite{HawEll73}. In the case of interacting fields, and 
after some 
algebra, we have arrived at the following expression,
valid for any spacetime obtained from
(\ref{G_general}): 
\begin{equation}
q=2\,\frac{c_1+c_2\,\dot G+c_3\dot G^2}{\left(c_4+c_5\,\dot G\right)^2}, \label{qgeneral}
\end{equation}
where
\begin{eqnarray}
&&c_1=4\,{p^2}\,\left[ {k^2}\,\left( {k^2} + 2\,n \right)  + 
6\,{n^2}\,\left( {k^2} + 4\,n \right) 
     \right],\label{eq:c1}\\
&&c_2=4\,|p|\,|k|\,\mathrm{sign}(p\,k)\left( {k^2} + 4\,n \right) 
\,{\sqrt{{k^4} - 4\,{n^2}}},\\
&&c_3={{\left( {k^2} - 2\,n \right) }^2}\,\left( {k^2} + 4\,n \right),\\
&&c_4=2\,|p|\,|k|\,\textrm{sign}(p\,k)\,{\sqrt{{k^2} + 2\,n}},\\
&&c_5={\sqrt{{k^2} - 2\,n}}\,\left( {k^2} + 4\,n \right).
\end{eqnarray}
The equivalent expression for the interacting case could be obtained
by making the replacements $\vert p\vert\rightarrow p=\vert \vec p\,\vert$,
$\vert k\vert\rightarrow k=\vert \vec k\vert$, $n\rightarrow 1$ and 
$\textrm{sign}(p\,k)\rightarrow\cos\lambda$, where
\begin{equation}
\cos\lambda=\frac{\vec k\cdot\vec m}{k\,m}=\frac{\vec k\cdot\vec p}{k\,p},
\end{equation}
and using
\begin{equation}
c_1=4\,{p^2}\,\left[ {k^2}\,\left( {k^2} + 2 \right)\cos^2\lambda  + 
6\left( {k^2} + 4 \right) 
     \right]
\end{equation}
instead of (\ref{eq:c1}).
Safely, the denominator of $q$ never vanishes for $t\ge 0$, as can be easily tested.

Our first step in this analysis is to investigate the late time behaviour of
$q$. We get
\begin{equation}
\lim_{t\rightarrow\infty}q=\frac{2\,k^2-4\,n} {k^2+4\,n},
\end{equation}
which is positive in the range of validity of our solutions, thus indicating
that our models do not inflate at late times. This is a very useful result indeed 
and will be used later.

Let us try to find out now whether there is a period of time for which $q$ becomes 
negative, recalling that this is  the condition for the existence of 
a period of accelerated inflation. A quick look at the expression for $q$ shows that 
$\textrm{sign}(p\,k)$ 
and $\cos\lambda <0$ are respectively  the
necessary conditions for inflation, since otherwise $q$ would be definite positive.
In general, the quantity $q$ will either vanish
at a couple of time instants which we will denote $t_-$ and $t_+$, or will never
become null at all.
Moreover, from the fact that at late times $q$ is necessarily positive
it follows that $q$ will be negative between those hypothetical $t_-$ and $t_+$.
The question to formulate is whether those instants exist and  whether the 
sign change of the deceleration factor at $t_+$, which ends inflation, occurs 
during the model's history $t\ge0$.

For our purposes it suffices to give the  implicit equations of $t_-$ and $t_+$,
in the interacting case these read
{
\setlength\arraycolsep{2pt}
\begin{eqnarray}
\dot G(t_{\pm})&=&
\frac{2 |p|}{{{\left( {k^2} - 2\,n \right) }^{{\frac{3}{2}}}}}
\Big(\pm\frac{
{\sqrt{6}}\,{\sqrt{{k^4}\,n - {k^2}\,\left({k^2} -2\right) \,{n^2} - 
      2\,{k^2}\,{n^3} + 8\,{n^4}}}}{\sqrt{k^2+4\,n}}-
\nonumber\\
&&\qquad\qquad\qquad |k| \textrm{sign}(pk)\sqrt{k^2+2\,n}\Big),\label{implicit}
\end{eqnarray}
}\noindent
which can be solved provided that
\begin{equation}
\frac{k^2}{n}+1\le\sqrt{\frac{9\,n-1}{n-1}}\label{cond_inf_1}\,.
\end{equation}
Typically, in multifield models and for a given $k$, a minimum number of fields will be needed
for inflation to occur in multifield models.  Note that
in the particular case $n=1$ there will be  inflation for any value of $k$. 
Let us assume now that we are dealing with a multifield case and that condition
 (\ref{cond_inf_1}) is fulfilled.

Clearly, inflation will occur if $t_+\!>0$ and, under the assumptions
made before, we can  conclude that there will be an epoch of accelerated
inflation if $\dot G(t_+)>\dot G(0)$. Remarkably, this condition follows
from (\ref{cond_inf_1}), so the latter is actually not only a necessary but also
a sufficient condition for the existence of inflation. With these results
at hand we conclude that our non-interacting models are examples 
of the assisted inflation phenomenon. As a matter of fact, in the cases in which 
the individual potentials are too steep inflation can still occur
thanks to the cooperation of the fields if there are enough of them. 

The  implicit equations of $t_-$ and $t_+$
in the interacting case are slightly different:
\begin{eqnarray}
\dot G(t_{\pm})=
\frac{2 p}{(k^2-2)^{\frac{3}{2}}}
\left(\pm\frac{\sqrt{6}\sqrt{8 - {k^2}\,\left( {k^2}+2
 \right)\sin^2\!\lambda}}{\sqrt{k^2+4}}\,-k\sqrt{k^2+2}\,\cos \lambda\right)
\end{eqnarray}
and they can be solved provided that
\begin{equation}
{k^2}+1\le\sqrt{\frac{9-\cos^2\lambda}{1-\cos^2\lambda}}\,.\label{cond_inf_2}
\end{equation}
By recalling that in this case $k^2=k_1^2+\dots+k_n^2$ we can see that
the more fields there are, the more difficult it becomes for  (\ref{cond_inf_2}) 
to be satisfied and the less likely inflation is. Note that in the particular
case $\cos \lambda=-1$ there will be inflation regardless of the value of
$k$. In general lines, the role played by the number of fields in also agreement
with existing results. The standard interpreted of this behaviour
is a direct consequence of the way the friction depends on the field population.  
Moreover,
if (\ref{cond_inf_2}) holds, then $G(t_+)>G(0)$ is fulfilled too, and 
we draw the conclusion that the fulfillment
(\ref{cond_inf_2}) is in fact a necessary and sufficient condition for inflation.

Finally, it is interesting to look at the late time degree of isotropy
of the spacetimes under discussion.
According to an intuitive criterion for a perfect fluid model 
to become isotropic at late times~\cite{Barrow}
 the quotient between
the shear and the expansion of its fluid must tend to zero on that
very limit. We use here the standard definition for the shear 
scalar of the fluid \cite{HawEll73},
namely
\begin{equation}
\sigma^2=\frac{1}{2}\sigma_{\alpha\beta}\sigma^{\alpha\beta}
\end{equation} 
where $\sigma_{\alpha\beta}$ is the shear tensor.
In general, in the non-interacting case one has for our geometries
\begin{equation}
\frac{\sigma}{\theta}=\frac{\dot f G-\dot G}{\sqrt{3}\,(\dot f G+2\dot G)}
\end{equation}
and at late times it becomes
\begin{equation}
\lim_{t\rightarrow\infty}
\frac{\sigma}{\theta}=\frac{k^2-2\,n}{\sqrt{3}\left(k^2 +4\,n\right)}.
\end{equation}
The latter limit is a monotonically increasing function 
of $k^2/n$. So the larger the number of non-interacting fields, the
more isotropy at late times. Conversely, using $n=1$ and
$k^2=k_1^2+\cdots+k_n^2$ we see that a larger number of interacting fields
will result asymptotically in a  more anisotropic spacetime.

\section{Conclusions}
Families of exact solutions to the set of Einstein-Klein-Gordon equations
for Bianchi $\textrm{VI}_0$ geometries with multiple scalar fields
with exponential potentials have been derived. The models obtained include 
cases with both interaction and non-interaction among the fields, and we have shown 
the asymptotic stability of them all. In particular this means that 
those spacetimes characterised by parameters  compatible 
with assisted inflation act as attractors of other possible 
solutions with the same values of those parameters. 

Performing a kinematic analysis based on exact, yet particular,
examples like the ones used here has a clear advantage in
comparison with studies just concerned with asymptotic features.
The severe limitation of not having any information about early
stages in the evolution of the model is overcome here, and therefore
we can formulate and answer the more natural question of whether inflation
exists soon after the beginning of times.

A thorough analysis of the conditions under which inflation occurs has yielded the 
result that in the interacting cases profusion in the number of fields
redounds to the  improbability of inflation; whereas in the cases without 
interaction all the contrary happens, this is, inflation is assisted. A discussion 
on the late time 
isotropisation of these spacetimes has been included as well.

In our opinion, it would be of interest to extend this work by 
performing an analogous in-depth  analysis about 
the conditions for inflation in the setup of other isotropic and anisotropic 
cosmologies.

\section{Acknowledgements}
We are grateful to R.~B. P\'erez-S\'aez for useful comments, 
to J.~A. Valiente-Kroon for discussions
and to J. Carot for bibliographic help. This work has been carried out 
thanks to the financial
support of Eusko Ikaskuntza, the University of the Basque Country 
(Research Project~UPV172.310-EB150/98 and  General Research
Grant~UPV172.310-G02/99)
and the Ministry of Education, Culture and Sport (Research Project~BFM2000-0018).


\end{document}